\documentclass[twocolumn, times]{aastex63} 
\usepackage{amsmath}

\newcommand{\kms}{\ifmmode {\rm km\ s}^{-1} \else km s$^{-1}$\ \fi}
\newcommand{\ergs}{\ifmmode {\rm erg\ s}^{-1} \else erg s$^{-1}$\ \fi}
\newcommand{\lb}{\ifmmode L_{\rm Bol} \else $L_{\rm Bol}$\ \fi}
\newcommand{\ledd}{\ifmmode L_{\rm Edd} \else $L_{\rm Edd}$\ \fi}
\newcommand{\lx}{\ifmmode L_{\rm 2-10keV} \else  $L_{\rm 2-10keV}$\ \fi}
\newcommand{\ha}{\hbox{H$\alpha$}}
\newcommand{\hb}{\hbox{H$\beta$}}

\newcommand{\mbh}{\ifmmode M_{\rm BH}  \else $M_{\rm BH}$\ \fi}
\newcommand{\lv}{\ifmmode \lambda L_{\lambda}(5100\AA) \else $\lambda L_{\lambda}(5100\AA)$\ \fi}
\newcommand{\lbol}{\ifmmode L_{\rm Bol} \else $L_{\rm Bol}$\ \fi}

\newcommand{\oii}{\hbox{[O\,{\sc ii}]}}
\newcommand{\nii}{\hbox{[N\,{\sc ii}]}}
\newcommand{\sii}{\hbox{[S\,{\sc ii}]}}
\newcommand{\oiii}{\hbox{[O\,{\sc iii}]}}

\newcommand{\hii}{\hbox{H\,{\sc ii}}}

\newcommand{\oh}{\ifmmode 12+ \log({\rm O/H}) \else 12+log(O/H) \fi}
\newcommand{\mdot}{\ifmmode \dot{m} \else \dot{m} \fi }
\newcommand{\llog}{\ifmmode {\rm log} \else {\rm log} \fi }

\submitjournal{ApJ}

\shorttitle{Local metallicity}
\shortauthors{Berzaf .B et al.}

\begin{document}

\title{THE LOCAL STAR FORMATION RATE SURFACE DENSITY AND METALLICITY RELATION FOR STAR-FORMING GALAXIES}

\correspondingauthor{Yulong Gao, Xu Kong, Berzaf Berhane Teklu}
\email{ylgao@mail.ustc.edu.cn, xkong@ustc.edu.cn, berzaf12@mail.ustc.edu.cn}

\author[0000-0001-5932-135X]{Berzaf Berhane Teklu}
\affiliation{CAS Key Laboratory for Research in Galaxies and Cosmology, Department of Astronomy, University of Science and Technology of China, Hefei 230026, China}
\affiliation{School of Astronomy and Space Science, University of Science and Technology of China, Hefei 230026, China}

\author[0000-0002-5973-694X]{Yulong GAO}
\affiliation{CAS Key Laboratory for Research in Galaxies and Cosmology, Department of Astronomy, University of Science and Technology of China, Hefei 230026, China}
\affiliation{School of Astronomy and Space Science, University of Science and Technology of China, Hefei 230026, China}
\affiliation{Institute of Astronomy, The University of Tokyo, Osawa 2-21-1, Mitaka, Tokyo 181-0015, Japan}

\author[0000-0002-7660-2273]{Xu Kong}
\affiliation{CAS Key Laboratory for Research in Galaxies and Cosmology, Department of Astronomy, University of Science and Technology of China, Hefei 230026, China}
\affiliation{School of Astronomy and Space Science, University of Science and Technology of China, Hefei 230026, China}

\author[0000-0001-8078-3428]{Zesen Lin}
\affiliation{CAS Key Laboratory for Research in Galaxies and Cosmology, Department of Astronomy, University of Science and Technology of China, Hefei 230026, China}
\affiliation{School of Astronomy and Space Science, University of Science and Technology of China, Hefei 230026, China}

\author[0000-0002-2384-3436]{Zhixiong Liang}
\affiliation{CAS Key Laboratory for Research in Galaxies and Cosmology, Department of Astronomy, University of Science and Technology of China, Hefei 230026, China}
\affiliation{School of Astronomy and Space Science, University of Science and Technology of China, Hefei 230026, China}

\begin{abstract}
We study the relations between gas-phase metallicity ($Z$), local stellar mass surface density ($\Sigma_*$), and the local star formation surface density ($\Sigma_{\rm SFR}$) in a sample of 1120 star-forming galaxies from the MaNGA survey. At fixed $\Sigma_{*}$ the local metallicity increases as decreasing of $\Sigma_{\rm SFR}$ or vice versa for metallicity calibrators of N2 and O3N2. Alternatively, at fixed $\Sigma_{\rm SFR}$ metallicity increases as increasing of $\Sigma_{*}$, but at high mass region, the trend is flatter. However, the dependence of metallicity on $\Sigma_{\rm SFR}$ is nearly disappeared for N2O2 and N2S2 calibrators.  We investigate the local metallicity against $\Sigma_{\rm SFR}$ with different metallicity calibrators, and find negative/positive correlations depending on the choice of the calibrator. We demonstrate that the O32 ratio (or ionization parameter) is probably dependent on star formation rate at fixed local stellar mass surface density. Additional, the shape of $\Sigma_*$ -- $Z$ -- $\Sigma_{\rm SFR}$ (FMR) depends on metallicity calibrator and stellar mass range. Since the large discrepancy between the empirical fitting-based (N2, O3N2) to electronic temperature metallicity and the photoionization model-dependent (N2O2, N2S2) metallicity calibrations, we conclude that the selection of metallicity calibration affects the existence of FMR on $\Sigma_{\rm SFR}$.
\end{abstract}


\section{Introduction}
\label{sec:intro}

Understanding the physical process of the interstellar medium (ISM) is a key for determining a complete picture of galaxy formation and evolution. In particular, metallicity over galaxies is one of the physical quantities to implement a hint regarding evolution. The metallicity could produce insights into regulating galaxy outflow properties \citep{Finlator2008,Lilly2013,Belfiore2016a}. The enrichment gas outflow and inflow decreased the gas-phase abundance inside a galaxy. The inflows dilute the metal content, while the outflows remove metals from the ISM. The galaxy metallicity will reduce if the outflow gas is enriching beyond the current ISM, either through the direct escape of metal-rich ejection from the supernova explosion (SNe) or through galactic winds with enriched  metals. 
   
The relation of mass-metallicity (MZR) was established by \cite{Lequeux1979}, which indicates that  the metallicity of galaxy increases as increasing of the stellar mass, this recognizes that the galactic outflows controlling the metal content of the interstellar medium. Since it was presented observationally with the aid of \cite{Tremonti2004}, who determined a tight relation spanning upon three orders of magnitude within the mass and a factor concerning about 0.1 dex into metallicity, using a large sample of star-forming galaxies from Sloan Digital Sky Survey (SDSS). Moreover, the MZR presents an entirely similar shape impartially concerning the oxygen abundance calibrator, including a clear trend for $M_{*} < 10^{10}M_\odot$, then pulling down to the asymptotic value for higher stellar masses (e.g., \citealt{Kewley2008}).

In particular, \cite{Ellison2008} had already shown the dependence of MZR on star formation rate. Alternatively, \cite{Lara-Lopez2010} and \cite{Mannucci2010} suggested the correlation between metallicity and star formation rate (SFR), observing that at a fixed mass, the lower metallicity galaxy shows higher SFR. However, both of the SFR and metallicity increase with increasing of galaxy stellar mass. Besides, numerous studies confirmed the M-$Z$-SFR (FMR) relation (e.g.,
\citealt{Mannucci2010,Hunt2012,Yates2012,Andrews2013,Salim2014,Wu2016,SanchezAlmeida2018,SanchezAlmeida2019}).

Furthermore, several studies also questioned the presence of FMR relation. \cite{Sanchez2013} obtained $\hii$ regions from the CALIFA data set \citep{Sanchez2012} but did not find the secondary relation with SFR. \cite{Moran2012} had shown that their data does not demonstrate this secondary relation; however, they suggested a secondary relationship to the gas fraction. Furthermore, \cite{Rosales-Ortega2012} found a correlation with specific SFR (sSFR) based on the equivalent width (EW) of \ha. The local MZR does not exhibit a secondary relation yet maintaining the primary link between stellar mass and SFR. \cite{Kashino2016} revealed that they could not determine the secondary dependence between SFR and MZR presented by \cite{Mannucci2010} for a single metallicity calibrator. More recently, \cite{Barrera-Ballesteros2017} used different metallicity calibrators; at any of the calibrations, they have not found a robust secondary trend of MZR with neither SFR nor sSFR. However, in a recent review about the FMR, \cite{Cresci2019} found the presence of FMR at fixed stellar mass by reanalyzing the data of CALIFA and SDSS-IV MaNGA.

In addition, the mass-metallicity relation is also mainly important for stellar evolution. Many studies \citep{Moran2012,Rosales-Ortega2012,Sanchez2013,Gao2018a} explored the correlation of the local metallicity and the local stellar mass density for local star-forming galaxies using the GASS survey \citep{Saintonge2011}, CALIFA survey \citep{Sanchez2012}, and the MaNGA survey \citep{Bundy2015}, respectively. Those authors confirmed the existence of correlation between local metallicity and local surface mass density. Alternatively, they noticed that $\hii$ regions with higher mass surface density are metal-richer than lower densities.

Moreover, the relation between SFR and ionization parameter are not well understood. \cite{Kewley2002} defined the ionization parameter using two-line ratio $\oiii/\oii$. Lately, \cite{Nakajima2014} presented a correlation between the ionization parameter and the global physical properties of galaxies and found that higher ionization parameters are determined in much less massive galaxies with low metallicity. In addition, the anti-correlation between the ionization parameter and metallicity of $\hii$ region was also presented by \cite{Dopita2006}. In particular, \cite{Kaasinen2018} find that higher ionization parameters toward higher sSFRs for their star-forming galaxies. High ionization parameter has been proposed as the result of high SFRs, contributing to a larger reservoir of ionizing photons (e.g., \citealt{Kewley2013a}).

Although, significant studies about the local MZR and FMR did not take the sensitivity of ionization parameter into account the metallicity and star formation rate. In this study, we perform a large spatially resolved samples from MaNGA IFU survey to investigate the relation of local stellar surface density, metallicity, and local star formation rate surface density ($\Sigma_{*}$ -- $Z$ -- $\Sigma_{\rm SFR}$). We also take a look at the effects of the ionization parameter with different metallicity diagnostics.

The paper is organized as follows. In Section \ref{sec:data}, we demonstrate the sample selection for star-forming galaxies from the MaNGA survey, fitting process, determination of the gas-phase oxygen abundances, stellar mass surface densities, and other physical parameters. In Section \ref{sec:results}, we investigate the distribution and relation of $\Sigma_*$ -- $Z$ -- $\Sigma_{\rm SFR}$, and the dependence of the ionization parameter with metallicity. We discuss the implication of our results of the fundamental metallicity relation (FMR) and  $\Sigma_{\rm SFR}$ -- $Z$ in Section \ref{sec:discussion}. Finally, we summarize our results in Section \ref{sec:summary}. Throughout this paper, we adopt a flat $\Lambda$CDM cosmology with $\Omega_\Lambda=0.7$, $\Omega_{\rm m}=0.3$, and $H_0=70$ km s$^{-1}$ Mpc$^{-1}$.

\section{Data}
\label{sec:data}

    \subsection{MaNGA Overview}
    \label{subsec:manga}
    
    The MaNGA survey is to examine 10,000 galaxies using an integral field spectroscopy unit (IFU), across 2700 $deg^{2}$ at local universe $z \sim$ 0.03 \citep{Bundy2015}. The wavelength ranges from 3600 to 10300 $\rm \mathrm \AA$ with spectral resolution  $R \sim$ 1400 to 2600. The standard effective spatial resolution of the extracted data cubes can be used to illustrate by Gaussian with an FWHM $\sim$ 2.5\arcsec, and the spaxel size of released datacubes is 0.5\arcsec. The MaNGA IFU includes 29 fiber bundles in the SDSS field of view. A further detailed description of MaNGA instrument structure can be found in \cite{Drory2015}. 
    
    In this study, we extract the MaNGA data from SDSS-IV Data Release 14 (DR14) \footnote{http://www.sdss.org/dr14/manga/manga-data/catalogs} \citep{Abolfathi2018} as the initial sample that contains 2812 galaxies.

    \subsection{Spectral Fitting and Emission-line Measurements}
    \label{subsec:spec-fitting}
    
    Firstly, we correct the Galactic extinction by applying the color excess $E(B-V)$ using a map of Milky Way \citep{Schlegel1998}. For this study we use the public \textsc{STARLIGHT} spectral fitting code (\citealt{CidFernandes2005}) to fit the observational continuum spectra with a series of single stellar population (SSP) models from \cite{Bruzual2003}, assuming a \cite{Chabrier2003} initial mass function (IMF) and adopting the attenuation law of \cite{Calzetti2000}. 
    
    The fluxes of strong emission lines (e.g., $\rm \oii\lambda3727$, $\hb$, $\oiii\lambda\lambda4959,5007$, $\ha$, $\nii \lambda 6583$ and $\sii\lambda\lambda$6717,6731) are fitted with the Gaussian profiles by applying an IDL package \textsc{mpfit} \citep{Markwardt2009}. Following the method of \cite{Ly2014}, we measure the signal-to-noise ratios (S/N) of these emission lines. Using Balmer decrement and assuming \cite{Calzetti2000} attenuation law, we correct the interstellar reddening. We presume the intrinsic flux ratio ($\rm \ha / \hb)_{0}=2.86$ under the case-B recombination.
    
    \subsection{Sample Selection}
    \label{subsec:sample}
    
    Our parent sample has 2812 galaxies from SDSS DR14, and we suppose to use the benchmark of NUV - r $<$ 4 to select star-forming galaxies (\citealt{Li2015}), then 1120 galaxies are selected. Our main goal is to select star-forming regions, we appear the spaxels have signal to noise ratio S/N($\ha$) $>$ 5, S/N($\hb$) $>$ 5, S/N($\oiii\lambda 3727) > 5$, S/N($\oiii\lambda\lambda 4959,5007$) $>$ 5, S/N($\nii \lambda 6583$) $>$ 3, with continuum S/N greater than 3 and the equivalent width of $\ha$ greater than 10 $\rm \mathrm \AA$. We use the \cite{Kauffmann2003} boundary line in the BPT diagram (\citealt{Baldwin1981,Kewley2001, Kauffmann2003}) to exclude the spaxels affected by the active galactic nucleus (AGNs). 
    
\begin{figure}
       \includegraphics[width=0.45\textwidth]{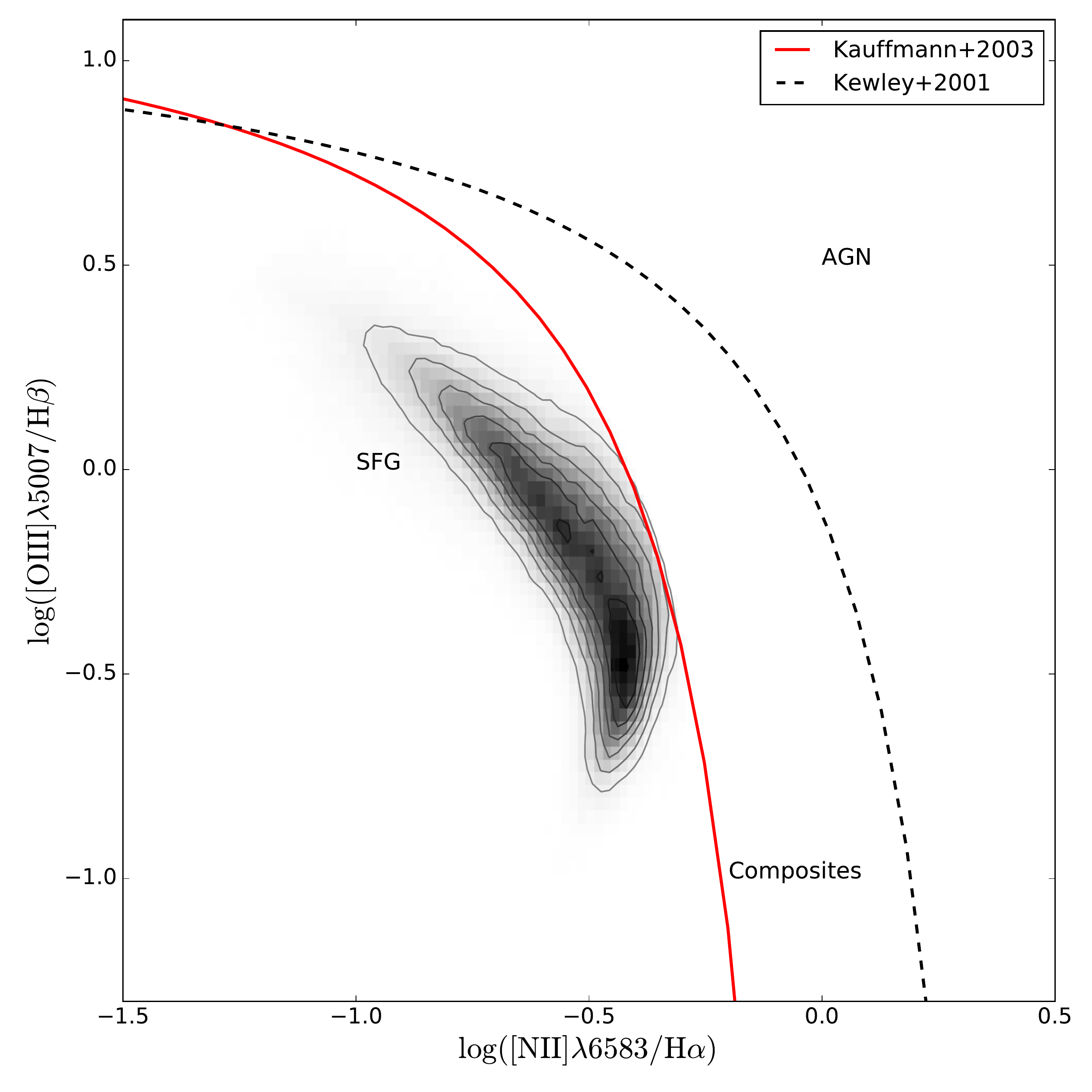}
        \caption{The BPT diagram for spaxels of star-forming galaxies in our sample. The grayscale histogram exhibit the observation number density. The red solid and black dashed lines represent the demarcation curves between SFGs and AGNs characterized by \cite{Kauffmann2003} and \cite{Kewley2001}, respectively.}
        \label{fig:bpt_manga}
    \end{figure}

    Figure \ref{fig:bpt_manga} illustrates the BPT diagram of about 740,000 spaxels. The grayscale exhibits the spaxel number density. As shown, our final sample of spaxels located in the star forming region.
    
    \subsection{Determinations of Metallicity, Star Formation Rate Surface Density, and Stellar Mass Surface Density}
    \label{subsec:metallicity}
    
    Determination of accurate oxygen abundance is more reliable using the electronic temperature or direct method, which can be obtained from the ratio of auroral to nebular intensity lines, such as \oiii$\lambda$4363/\oiii$\lambda$5007. Getting the oxygen abundance using the direct method is challengeable for higher metallicity because \oiii$\lambda$4363 becomes fainter. In this study, we will adopt four metallicity calibrations based on strong lines.
 
    I). The N2O2 calibrator is a more reliable estimator for local galaxies because of the spectral wavelength coverage. For nearby galaxies, we can get the lines of $\oii$$\lambda3727$ and $\nii$$\lambda6583$, but for the higher redshift, it is challengeable to observe those lines. Additional, the ratio of $\nii$ and $\oii$ is less affected by the ionization distribution, thus the N2O2 can be an oxygen abundance indicator. The calibrated metallicity using the relation in \cite{Kewley2002} is given by:
    \begin{equation}
    \begin{split}
    \rm 12 + log(O/H) = log(1.54020 + 1.26602 \times N2O2 \\
     \rm + 0.167977 \times N2O2^2)+ 8.93,
    \end{split}
    \end{equation}
    where $\rm N2O2 \equiv log[(\nii\lambda6583)/(\oii\lambda\lambda3727,3729)]$, with estimated uncertainty of 0.04 dex.
    
    II). O3N2: This index is commonly known as that it depends on strong emission line flux ($\ha$, $\hb$, $\oiii\lambda5007$, $\nii\lambda 6583$) to diagnose oxygen abundances in the literature.  The diagnostic of O3N2 index \citep{Alloin1979} is characterized by
    \begin{equation}
    \rm O3N2 \equiv log(\frac{\oiii\lambda5007}{\hb} \times \frac{\ha}{\nii\lambda6583}).
    \end{equation}    
    \cite{Pettini2004} recalibrate O3N2 diagnostic with $T_{e}$ based metallicity for $\hii$ region samples. Later, \cite{Marino2013} improved the O3N2 calibration with a large sample of $\hii$ regions. Here we use the modified O3N2 calibration by \cite{Marino2013}, which is given by
    \begin{equation}
    \rm 12 + log(O/H) = 8.505 - 0.221 \times O3N2.  
    \end{equation}
    
    III). N2:  \cite{Storchi-Bergmann1994} initially suggested the N2 index as a calibrator because the N2 index strongly correlated with abundance. The diagnostic N2 index \citep{Storchi-Bergmann1994,Raimann2000} is described as    
     \begin{equation}
    \rm N2 \equiv log(\nii\lambda6583 / \ha).
    \end{equation}
    Hence, N2 calibrated metallicity is determined by \cite{Marino2013} and expressed as  
    \begin{equation}
    \rm 12 + log(O/H) = 8.667 + 0.455 \times N2.
    \end{equation} 
    
    IV). N2S2 index: The $\nii\lambda$6583/$\sii\lambda\lambda$6717,6731 diagnostic is sensitive to metallicity and weakly dependent on the ionization parameter.  The N2S2 index is defined as:
    \begin{equation}
    \rm N2S2 \equiv log(\frac{\nii\lambda6583}{\sii\lambda\lambda6717,6731}), 
    \end{equation}
    and the metallicity calibration relation using the N2 and N2S2 from \cite{Dopita2016} is given as
    \begin{equation}
    \rm 12 + log(O/H) = 8.77 + N2S2 + 0.264 \times N2.
    \end{equation} 
    
    The global stellar mass $M_{*}$ and minor-to-major axis ratio ($b/a$) are extracted from the NSA catalog \footnote{http://www.nsatlas.org} (\citealt{Blanton2005}, \citealt{Blanton2011}). Following the method of \cite{Barrera-Ballesteros2016}, we generate the local stellar mass surface density $\Sigma_*$. We divide the surface mass density in each spaxel from the output results of \textsc{STARLIGHT} by its corresponding physical area and then correct the inclination using the $b/a$.
  
     We use the dust-corrected \ha \ emission-line luminosity to determine the star formation rate for each spaxel, using  the formula from \citet{Kennicutt1998} and assuming a \cite{Chabrier2003} IMF.

\section{Result}
\label{sec:results}
In this section, we utilize the dependence of $\Sigma_{*}$ -- $Z$ relation as a function of the local $\Sigma_{\rm SFR}$. Moreover, we explore the dependence of the $\Sigma_*$ -- $Z$ relation on star formation rate surface density and the ratio of $\oiii\lambda5007/\oii\lambda3727$.

    \subsection{The Dependence of the $\Sigma_{*}$ -- $Z$ Relation on Star Formation Rate Surface Density} 
    \label{subsec:metallicity-sfr}
    
    In \cite{Gao2018a}, we reported the relationship between the local stellar mass surface density and gas metallicity and found similar relations with \cite{Rosales-Ortega2012}and \cite{Sanchez2013}. As proposed in these results, it is necessary to check if the local $\Sigma_{*}$ -- $Z$ relationships have a secondary relationship with local $\Sigma_{\rm SFR}$. 
    
\begin{figure*}
      \includegraphics[width=1.\textwidth]{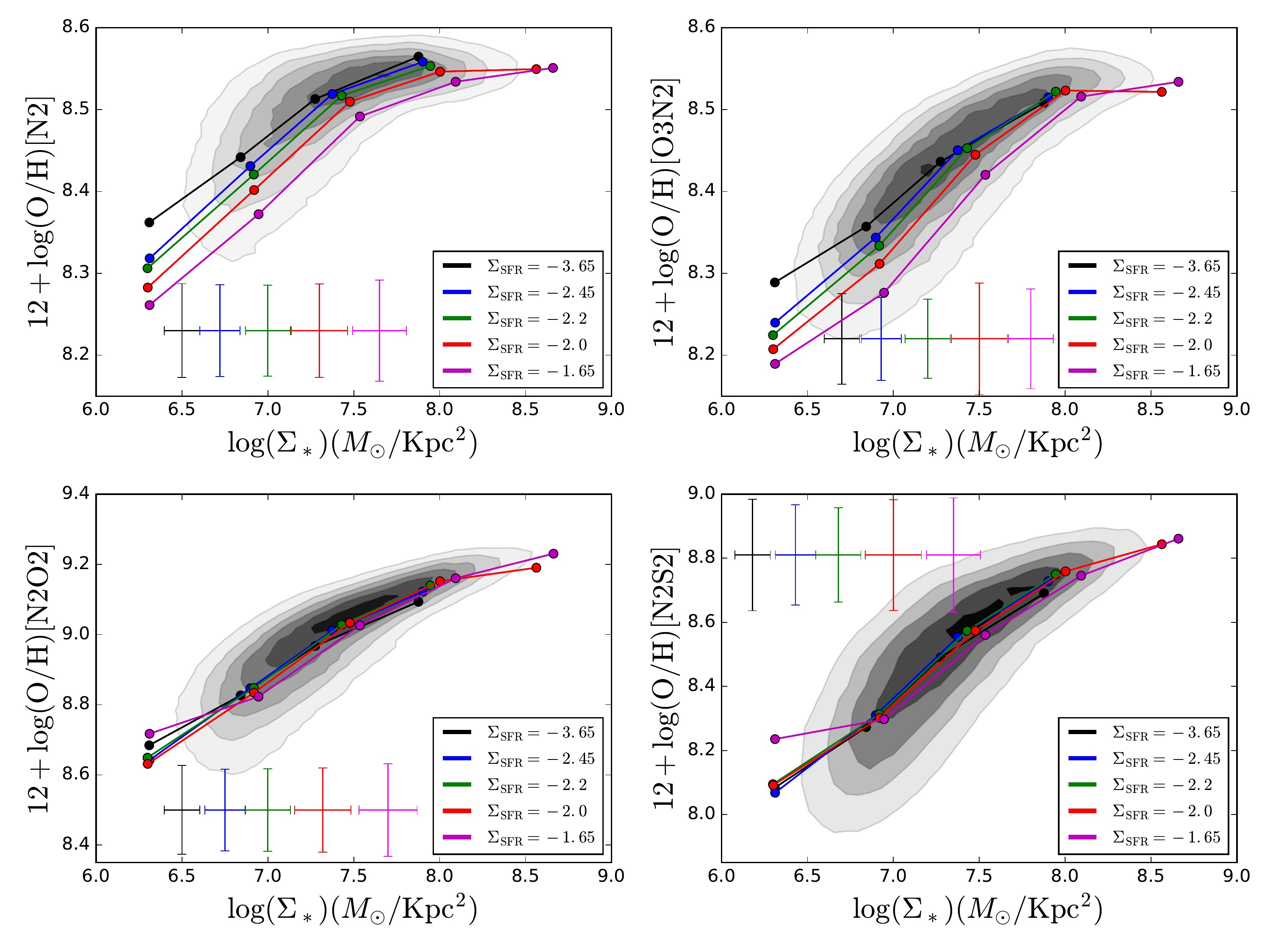}
      \caption{The distribution of metallicity against local stellar mass surface density as regards four different metallicity calibrations, O3N2, N2, N2O2 and N2S2 indexes. We separate our spatially resolved MaNGA samples into five subsamples based on local star formation surface densities ($\Sigma_{\rm SFR}$), which are binned as $-3.69, -2.45, -2.2, -2$ and $-1.65$. The left top and right top panels shows the relation between $\Sigma_{*}$ -- $Z$ based on N2 and O3N2 metallicity indicators, respectively. The bottom left and right panels indicates $\Sigma_{*}$ -- $Z$ relation using the N2O2 index and N2S2 index, respectively. The color-coded symbols with linked lines indicate the median values for different $\Sigma_{\rm SFR}$ bins. The error-bars serve as the standard deviation of median values in different bins.}
      \label{fig:mass_z_sfr}
    \end{figure*}

    As mentioned above, we investigate the dependence of local metallicity and the local stellar mass surface density $\Sigma_{*}$ as a function of local star formation rate surface density ($\Sigma_*$ -- $Z$ -- $\Sigma_{\rm SFR}$) relation for our samples in Figure \ref{fig:mass_z_sfr}. We separate our sample into five different bins based on local star formation rate surface density, and each panel shows different metallicity indicators. The bins are ranging from lower to higher $\Sigma_{\rm SFR}$ $(-3.65, -2.45, -2.2, -2$ and $-1.65)$ values. The median values for different star formation ranges are shown as the different color code connected lines.
    
    In the top left and right panels of Figure \ref{fig:mass_z_sfr}, we present the metallicity derived by N2 and O3N2 indexes as a function of the local stellar mass surface density. As shown in these panels, the metallicity increases with increasing of local stellar mass surface density. Compared to the higher star formation rate surface density region, the lower star formation rate surface density region shows higher metallicity at a fixed $\Sigma_{*}$. In this case, there is a tendency that the star formation rate surface density decreases as the metallicity increases.
    
    The bottom left and right panels of Figure \ref{fig:mass_z_sfr}, show the metallicity derived with N2O2 and N2S2 indexes, respectively. In these panels, the trend of $\Sigma_{*}$ -- $Z$ is similar to N2 and O3N2, however, not obvious as a function of local $\Sigma_{\rm SFR}$. Interestingly, we notice that the $\Sigma_{\rm SFR} $ does not change with increasing metallicity at fixed $\Sigma_{*}$. In particular, the metallicity increases steeply with local stellar mass surface density, while flatter relation appears at high mass surface density (${\rm log}(\Sigma_*) >  7.8$) for the top two panels. However, the difference in the dependence of local MZR on $\Sigma_{\rm SFR}$ is attributable to the metallicity calculation method between different diagnostics and calibrations, more detail about this discrepancy can be found in previous studies \citep[e.g.,][]{Pettini2004,Kewley2008,Sanchez2012}. On the other hand, the $\ha$ based estimation of $\Sigma_{\rm SFR}$ also affects the shape of $\Sigma_{*}$ -- $Z$ -- $\Sigma_{\rm SFR}$. We will show more detail about this reversal relation for those different metallicity indicators in the following sections.

    \subsection{Star Formation Rate Surface Density and Metallicity ($\Sigma_{\rm SFR}$ -- $Z$) Relation on $M_{*}$}
    \label{subsec:metallicity-mass-sfr}
    
    In the previous section, we examine the relation between $\Sigma_{*}$ -- $Z$, and $\Sigma_{\rm SFR}$. In this section, we aim further to test the dependence of $\Sigma_{\rm SFR}$ on the local metallicity comparing different metallicity calibrations at fixed stellar mass and local stellar mass surface density. To quantify this relation, for each subsample, we derive the correlation coefficient for checking if it presents any variation with the stellar mass. If the $\Sigma_{\rm SFR}$ -- $Z$ change significantly for each stellar mass bin, maybe the shape of $\Sigma_{*}$ -- $Z$ truly depends on the $\Sigma_{\rm SFR}$, as we have shown in the top left and right panels in Figure \ref{fig:mass_z_sfr} of Section \ref{subsec:metallicity-sfr}.
    
     \begin{figure*}
     \includegraphics[width=1.\textwidth]{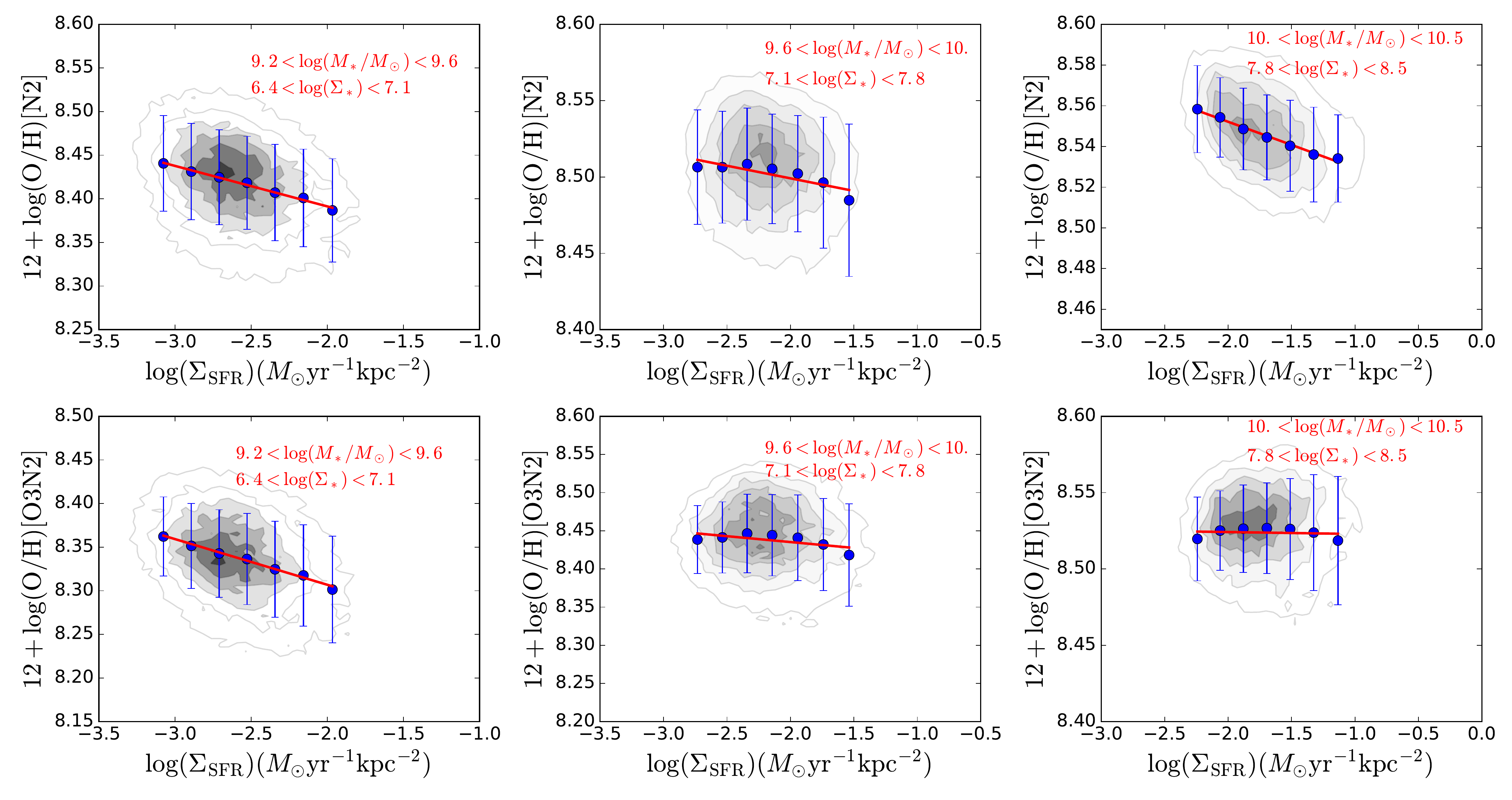}
     \includegraphics[width=1.\textwidth]{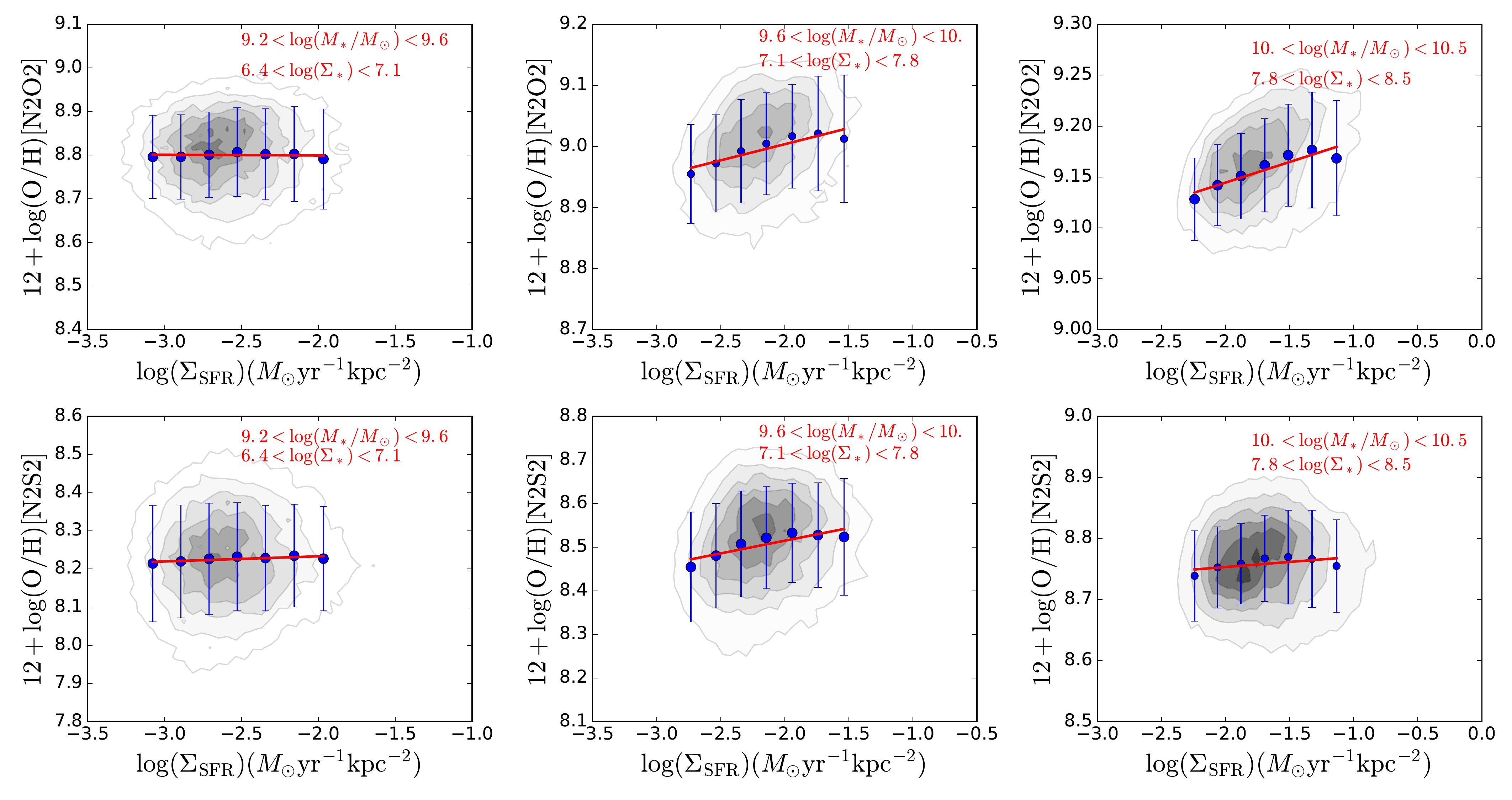}
      \caption{The metallicity distribution in $\Sigma_{\rm SFR}$ space based on O3N2, N2, N2O2, N2S2 indexes. We separate the plotting range $M_*$ and $\Sigma_*$ into three bins ($M_*: 9.2, 9.6, 10.0, 10.5$ \& $\Sigma_*: 6.4, 7.1, 7.8, 8.5$). The left four rows show the lower mass bin for different metallicity calibration; middle four panels represent the intermediate mass range and the last right four panels show for a higher mass bin for different metallicity calibration, respectively. The contour represents the distribution of $\Sigma_{\rm SFR}$ -- $Z$. The red dashed line in each panel shows the best-fitted $\Sigma_{\rm SFR}$ -- $Z$ relation of subsamples. Blue symbols and error-bars display as the median values and standard deviation, respectively.}
      \label{fig:plot_mz_all}
      \end{figure*}
    
    In Figure \ref{fig:plot_mz_all}, we show the distribution of metallicity-star formation rate surface density, at fixed global stellar mass and local stellar mass surface density regardless of \cite{Gao2018a}. The first column shows the distribution at low mass range of $9.2 < {\rm log}(M_{*}/M_\odot) < 9.6$ and $6.4 < {\rm log}(\Sigma_{*}) < 7.1$  with different metallicity calibrations. Second column represents the intermediate-mass bin $9.6 < {\rm log}(M_{*}/M_\odot) < 10.0$ and $7.1 <  {\rm log}(\Sigma_{*})  < 7.8$. The last column indicates the high mass bin $10.0 < {\rm log}(M_{*}/M_\odot) < 10.5$ and $7.8 < {\rm log}(\Sigma_{*}) < 8.5$. In each panel, the solid red line indicates the best-fit liner relation of $\Sigma_{\rm SFR}$ -- $Z$, and the gray contour represents scatter of subsamples.

\begin{table}[ht]
\begin{center}
\caption{Correlation Coefficient ($r$ values) of N2, O3N2, N2O2 and N2S2 Metallicity Indices.\label{table1}}
\begin{tabular}{@{}lrrrrr@{}}
\tableline
\tableline
${log}(M_{*}/M_\odot)$                          & N2              & O3N2    & N2O2          & N2S2        \\ 
 
 \tableline                               
9.2 -- 9.6                      &   -0.33           & -0.386        & 0.009           & 0.056            \\
9.6 -- 10.                     &   -0.256           & -0.203       & 0.132           & 0.092           \\
10. -- 10.5                      &  -0.279           & -0.025       & 0.269           & 0.127             \\
\tableline
\tableline
\end{tabular}
\begin{flushleft}
Notes: We separate out the sample according to global stellar mass and local stellar mass surface density. The mass range of first row is $9.2 < {\rm log}(M_*/M_\odot) < 9.6$ and $6.4 <  {\rm log}(\Sigma_{*}) < 7.1$. Second row mass bin $9.6 < {\rm log}(M_*/M_\odot) < 10.$ and $7.1 <  {\rm log}(\Sigma_{*}) < 7.8$. The last row shows the range of $10. < {\rm log}(M_*/M_\odot) < 10.5$ and $7.8 <  {\rm log}(\Sigma_{*}) < 8.5$. 
\end{flushleft}
\end{center}\end{table}
    
    For low mass bin subsamples, we find a clear and strong anti-correlation between $\Sigma_{\rm SFR}$ and metallicity, with $r = -0.33$ and $r = -0.38$, for metallicity index N2 and O3N2, respectively. However, for N2O2 and N2S2 metallicity indices, no correlation is found at this mass range. Our results regarding correlation coefficients of different metallicities with bins of stellar mass and local surface mass density are shown in Table \ref{table1}. 

    As shown in Figure \ref{fig:plot_mz_all}, the $\Sigma_{\rm SFR}$ -- $Z$ relation at different stellar mass bins shows different  slopes for N2, O3N2, N2O2, and N2S2 metallicity calibrators. The trends are slightly similar for those two calibrators of N2 and O3N2. If we focus on the low and intermediate-mass bins, we can see that the $\Sigma_{\rm SFR}$ -- $Z$ slope presents anti-correlation. This trend is consistent with \cite{SanchezAlmeida2019} at stellar mass range $9.2 < {\rm log}(M_*/M_\odot) < 9.6$ using O3N2 index. Besides considering the N2 index, at fixed mass, the metallicity decreases with increasing of local $\Sigma_{\rm SFR}$.

    As we explored, the relations indicate that for higher (lower) stellar mass bins, the correlation is weak (strong) for $\Sigma_{\rm SFR}$ -- $Z$. As we shown, for O3N2 metallicity calibration, at higher stellar mass (${\rm log}(M_{*}/M_\odot) > 10.$), we find a flatter trend, which is similar to the flat MZR at high mass (e.g., \citealt{Tremonti2004,Kewley2008,Zahid2014}).  However, Figure \ref{fig:plot_mz_all} shows a positive correlation for N2O2 and N2S2 calibrators, while a negative correlation for N2 index, at high mass region.
    
    Generally, this mixed type of behavior is different from the local MZR and FMR at different stellar mass bin regions for star-forming galaxies reported in other studies.
    The mixed results on this $\Sigma_{\rm SFR}$ -- $Z$ relations are possibly caused by different metallicity estimation, which have systematic errors on calibration, and the calculation of $\Sigma_{\rm SFR}$ from strong emission line. The variations among observed behavior are just probably associated along with the methods used to measure the oxygen abundance among different studies.
    
    We expect that the different trends in $\Sigma_{\rm SFR}$ -- $Z$ relations for different stellar mass bins, with four metallicity calibrators, might be caused by the dependence of metallicity on ionization parameter. We will discuss the ionization parameter in the following sections. 

    \subsection{Dependence of $\oiii\lambda5007/\oii\lambda3727$ on Metallicity}
    \label{subsec:o32-metallicity}
\begin{figure*}
     \includegraphics[width=1.\textwidth]{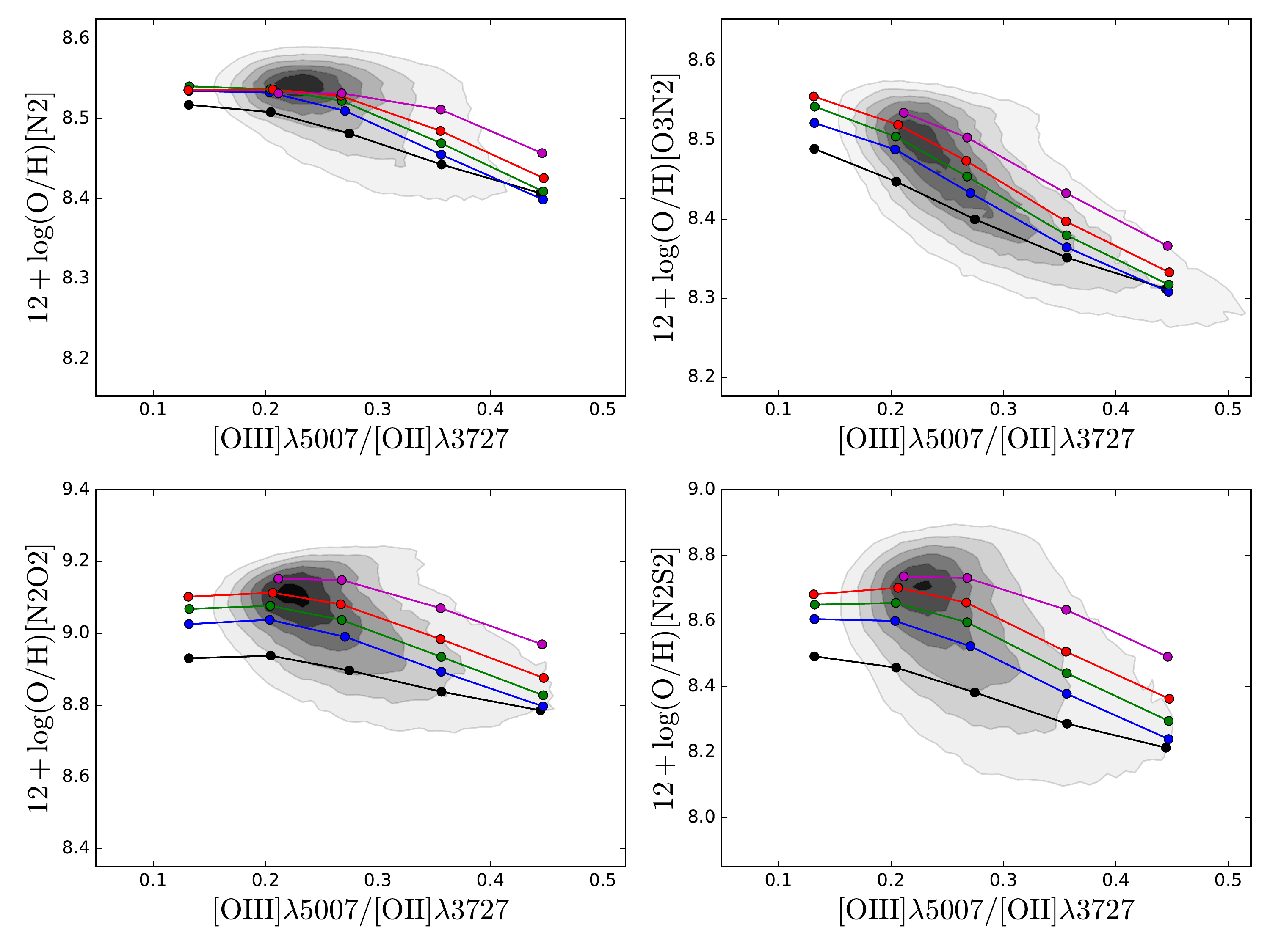}
    \caption{The median metallicity computed with four different metallicity calibrations, O3N2, N2, N2O2, and N2S2 index as a function of the O32 ratio. We separate our spatially resolved MaNGA samples based on local star formation surface density ($\Sigma_{\rm SFR}$) in five subsamples. The color-code and symbols are same as Figure \ref{fig:mass_z_sfr}.}
    \label{fig:plot_ion_z}
    \end{figure*}

    In this section, we demonstrate the dependence of $\oiii\lambda5007/\oii\lambda3727$ on the metallicity. The ratio of O32 is a proxy of the ionization parameter \citep{Dopita2000,Kewley2002}. In Figure \ref{fig:plot_ion_z}, we show the metallicity distribution as a function of the O32 ratio with four different calibrators. Different color means different $\Sigma_{\rm SFR}$, which are same as Figure \ref{fig:mass_z_sfr}. The median metallicity is shown for bins of varying $\Sigma_{\rm SFR}$. In particular, at a fixed ratio of O32, the metallicity increases with the increasing of $\Sigma_{\rm SFR}$. 
    
    In Figure \ref{fig:plot_ion_z} we see that the local metallicity is decreasing with the increasing of O32 ratio. As shown in Figure \ref{fig:plot_ion_z}, the higher metallicity corresponds with a lower O32 ratio, which is also described by \cite{Maier2004}. \cite{Dopita2006} argued that the higher opacity stellar wind causes a strong dependence between chemical abundance and ionization parameter in the surrounding $\hii$ region. We investigate that the metallicity calibrator depends on the ionization parameter or the ratio of O32. We expect that the existence of anti-correlation between the ionization parameter and metallicity can interpret the presence of anti-correlation between $\Sigma_{\rm SFR}$ and metallicity.

    \subsection{The Relation Between Ratio of O32, $\Sigma_*$ and $\Sigma_{\rm SFR}$}
    \label{subsec:o32-mass-sfr}

    \begin{figure}
    \includegraphics[width=0.45\textwidth]{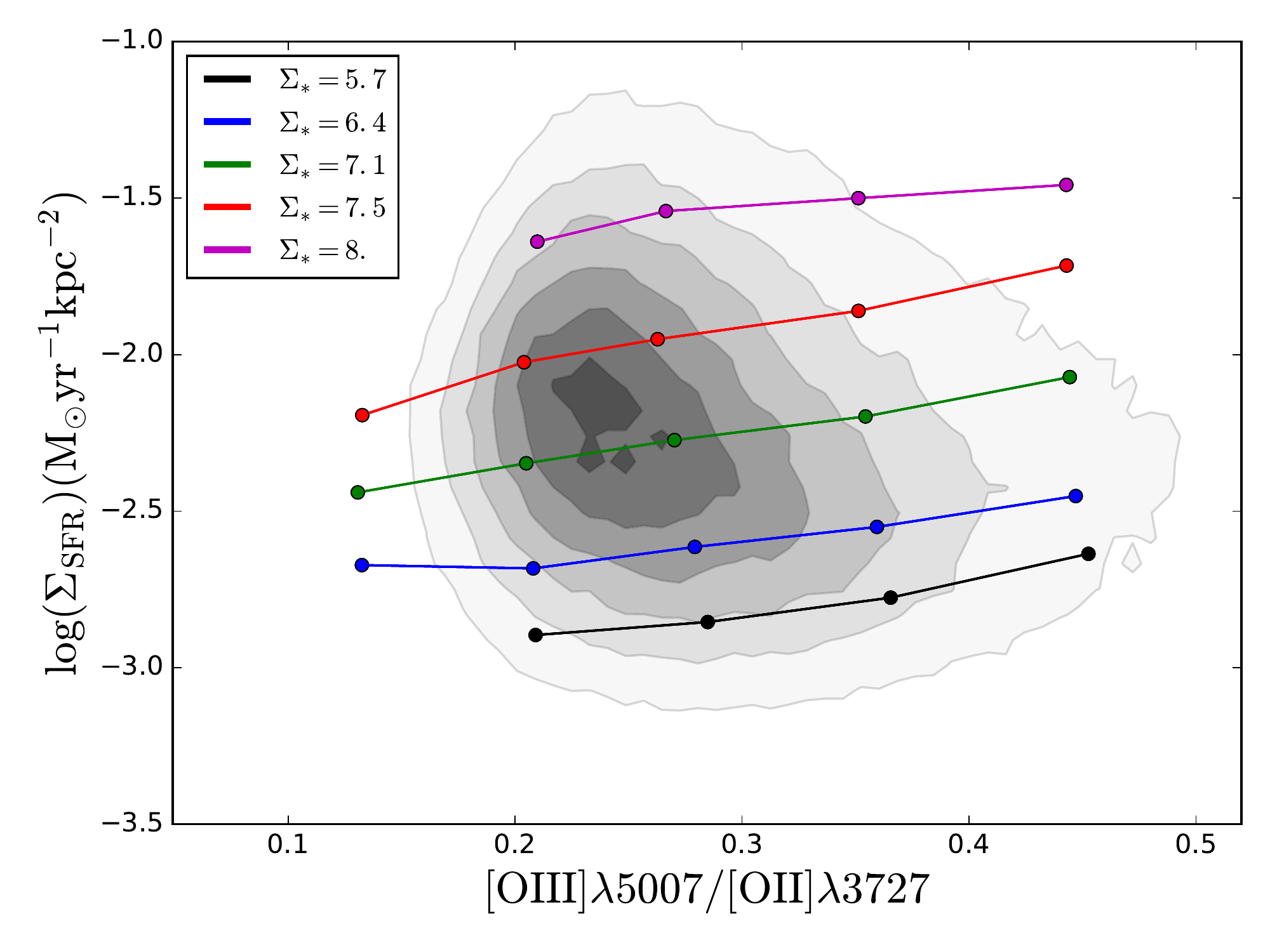}
    \caption{The relation of O32 verses local star formation rate surface density as function of local stellar mass surface density. The different colors indicate the bins of local stellar mass surface density range (${\rm log}(\Sigma_{*})$: $5.7, 6.4, 7.1, 7.5$ and $8.$)}       
    \label{fig:plot_sfr_q_all}
    \end{figure}

    We now demonstrate the correlation between the ionization parameter, $\Sigma_*$, and $\Sigma_{\rm SFR}$. In Figure \ref{fig:plot_sfr_q_all}, the different colors represent the bins in different ranges of local stellar mass surface density. The connected lines indicate the median values of the ratio of O32 and $\Sigma_{\rm SFR}$ in each bin of $\Sigma_*$.
 
    The $\Sigma_{\rm SFR}$ weakly increases with increasing O32, while at the fixed ionization parameter, the regions with higher $\Sigma_*$ have higher $\Sigma_{\rm SFR}$ simultaneously. We find a weak positive correlation between the O32, $\Sigma_*$, and $\Sigma_{\rm SFR}$ relation, which is in agreement with \cite{Kewley2015}. We support the scenario that, a more significant number of stars within $\hii$ regions, are probably to be mainly accountable for varying the ionization parameter with stellar mass surface density. 
    
    \subsection{The Fundamental Metallicity Relation}
    \label{subsec:fmr}
    As we mentioned  in Figure \ref{fig:mass_z_sfr}, the $\Sigma_{*}$ -- $Z$ as a function of $\Sigma_{\rm SFR}$, for the upper left and right panels, we demonstrate a clear trend with these two metallicity indicators. Here in this section, we suppose to check the secondary relation of $\Sigma_{\rm SFR}$.
    \cite{Mannucci2010} present the existence of the FMR, who estimated the scatter by adopting median metallicity values on the projection MZR space. The projection is described as :
    \begin{equation}
    \mu_{\alpha}= {\rm log}(\Sigma_*) - \alpha {\rm log}(\Sigma_{\rm SFR}).
    \end{equation}

    Interestingly, except at lower stellar mass and higher specific star formation rate, \cite{Barrera-Ballesteros2016} presented that the relationship of $\Sigma_{*}$ -- $Z$ is broadly independent on the total stellar mass of galaxy and specific star formation rate, using O3N2 metallicity calibrator. Before applying FMR projection relation, we excluded the galaxies with lower stellar mass, by setting ${\rm log}(M_*/M_\odot) > 9.6$, and we use $\alpha=0.32$ as proposed by \cite{Mannucci2010} to reduce the scatter of our sample. 

\begin{figure*}
    \begin{center}
    \includegraphics[width=1.\textwidth]{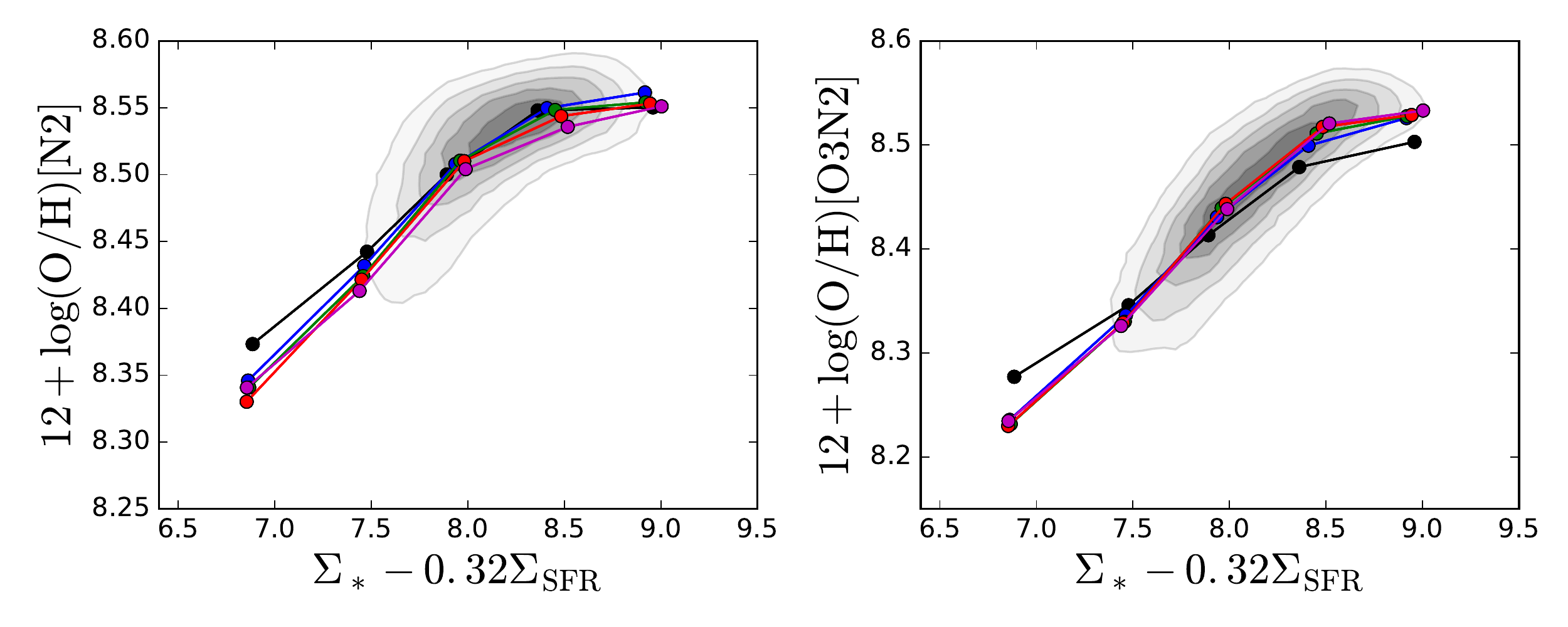}
    \caption{The local metallicity distribution as a function of FMR projection $\mu_{\alpha}$. The connected color coded lines and gray contour are same as Figure 2.  The left and right panels represent FMR for O3N2 and N2 metallicity calibrators.}
    \label{fig:plot_fmr}
    \end{center}
    \end{figure*}

    In Figure \ref{fig:plot_fmr}, we display the local metallicity distribution as a function of FMR projection $\mu_{\alpha}$, consistent with \cite{Mannucci2010}. We find the scatters of our MaNGA sample are $\sigma = $ 0.28 dex and 0.26 dex using metallicity calibrations of N2 and O3N2, respectively. We confirm the existence of the trend in Figure. \ref{fig:mass_z_sfr} that regions with  higher $\Sigma_{\rm SFR}$ have lower metallicity at fixed $\Sigma_*$.  However, the $\alpha$ might not be fitted for the lower $\Sigma_{\rm SFR}$ region (black lines in Figure \ref{fig:plot_fmr}). 

\section{Discussion}
\label{sec:discussion}
In this study, we demonstrate the local $\Sigma_*$ -- $Z$ relation and their dependence on the $\Sigma_{\rm SFR}$ and ionization parameter using different metallicity calibrations for spatially resolved samples about 740,000 spaxels from MaNGA survey. 

We confirm the tight $\Sigma_*$ -- $Z$ relation reported by \cite{Rosales-Ortega2012} and \cite{Barrera-Ballesteros2016}. Furthermore, we explore the effect of $\Sigma_{\rm SFR}$ on the local surface mass density and local metallicity relation or the fundamental relation. We found a clear correlation between $\Sigma_*$ -- $Z$, and $\Sigma_{\rm SFR}$ for the metallicity calibration N2 and O3N2 indexes. Our results in Figure \ref{fig:mass_z_sfr} is consistent with global and local FMR studies (e.g., \citealt{Lara-Lopez2010,Mannucci2010,Salim2014,Sanchez2013,Sanchez2017,Sanchez2019}) only with N2 and O3N2 metallicity calibrations. 

    \subsection{Physical Explanation of O32 Ratio and Local Metallicity}
    \label{subsec:explanation}
    The metallicity anti-correlates with the ratio of O32, which causes the discrepancy with empirical emission lines. \cite{Ho2015} presented the comparison of different diagnosis, and found that the diagnosis of O3N2 offers higher metallicity than the N2O2 at a higher ionization parameter or vise-verse.

    The ionization parameter and O32 ratio commonly demonstrate dependence on metallicity since the ionizing spectrum related to metallicity \citep{Kewley2002}. Although low-metallicity stars generate higher ionizing photons and have a harder ionizing spectrum \citep{Leitherer2014} as that of an anti-correlation with metallicity and ionization parameter. With these caveats in mind, we present the evidence of the anti-correlation of the metallicity and O32, as shown in Figure \ref{fig:plot_ion_z}. Furthermore, \cite{Sanders2016} found that the hardness of the ionizing spectrum increases with decreasing metallicity and higher metallicity objects to have lower O32, lower metallicity tends to higher O32 values.
    
    In other word, as we examine the lower/higher ionization parameter may vary because of the fluctuating the hardness of the ionizing spectrum with metallicity. Regarding this, \cite{Nagao2006} proposed two possibilities, the first one is due to mass-metallicity and mass-age relations. According to this, more massive and older systems are linked with a higher metallicity of galaxies. Due to the decreased luminosity of the ionizing stars, $\hii$ regions ionized by later stellar populations are supposed to be distinguished by lower ionization parameters. The other possibility is, the relationship between gas metallicity and stellar metallicity, because lower metallicity stars ionize lower metallicity gas. The first possibility can be checked by the O32 ratio with $\Sigma_{\rm SFR}$ as shown in Figure \ref{fig:plot_sfr_q_all}. Indeed, this related to at a fixed mass surface density, there is an increasing ionization parameter with $\Sigma_{\rm SFR}$, this may appear due to the fraction of young stars to old stars in a galaxy.
   
    In addition, as we have seen in Figure \ref{fig:mass_z_sfr}, the results of different metallicity calibrations agree with suggestions of \cite{Curti2017} and \cite{Curti2019}, they found the $T_{e}$ method metallicity calibration have lower metallicity value than photoionization model calibrations. Indeed, this indicates the higher values of N2O2 and N2S2 are reasonable compared to N2 and O3N2 calibrations. The main reason is that this difference might be caused by the bias at the higher excitation conditions.

  \subsection{The FMR and  $\Sigma_{\rm SFR}$ -- $Z$}
  \label{subsec:fmr-z-sfr}
    Using the metallicity calibration N2 and O3N2, there is an anti-correlation between metallicity and  $\Sigma_{\rm SFR}$, which recover as \cite{Mannucci2010}, however, it changes to positive correlation at the high mass when we use N2O2 and N2S2 indexes. During this condition, did the FMR exist?
 
    \cite{Yates2012} presented the FMR projection, while the SFR anti-correlates to metallicity at lower stellar masses, then the relationship inverts to a positive correlation at higher stellar masses. They argued that the inversion is the consequence of gas-rich mergers at higher stellar masses fueling a starburst. Later, \cite{Lara-Lopez2013} found a similar relation at higher mass; it gives the reversal of the FMR, higher SFR exhibits higher metallicity. However, \cite{Telford2016} presented the systematic effects of the secondary dependence of the MZR on SFR and had a comparison with \cite{Mannucci2010}, suggested that it is weaker for the secondary relationship. 

    Results in Figure \ref{fig:plot_mz_all}, have shown a mixed scenario of negative and positive correlations. For N2 and O3N2 metallicity calibrations at low and intermediate mass, we found the negative $\Sigma_{\rm SFR}$ -- $Z$, which is consistent with \cite{Mannucci2010}, suggesting that the lower mass produces lower metallicity and higher $\Sigma_{\rm SFR}$. On the other hand, this analysis supports the original FMR, only if we use N2 and O3N2 indexes. Indeed, \cite{Lian2015} found at fixed stellar mass, the lower metallicity corresponds to a younger stellar population for local Lyman-break analogs (LBAs) galaxies. For the N2O2 and N2S2 metallicity indicators, the result is challengeable to interpret, as we have seen the positive trend for high mass range.  This inverted result is possibly caused by the metallicity calibration. In Section \ref{sec:results}, we have seen that for higher $\Sigma_{*}$ using N2O2 and N2S2, the MZR is increasing for higher $\Sigma_{\rm SFR}$, but is flatter for high mass with N2 and O3N2 calibrators. 
    
    Looking back to the two bottom panels of Figure \ref{fig:mass_z_sfr}, the local metallicity increases with increasing local stellar mass surface density but no trend with local SFR. In Figure \ref{fig:plot_mz_all}, the $\Sigma_{\rm SFR}$ -- $Z$ increases with the increasing stellar mass, this indicates that the stellar mass and local stellar mass surface density are physically fundamental to drive the local metallicity in $\hii$ regions.

    \cite{Kashino2016} reported that for local galaxies at low mass, the median value of metallicity shows flat or constant. They suggest that this result is due to the dominance of the primary production of nitrogen in less massive galaxies. Their finding is in agreement with our results using the metallicity indicators of N2O2 and N2S2 in Figure \ref{fig:plot_mz_all}, the $\Sigma_{\rm SFR}$ -- $Z$ relation is constant at the lower mass bins. Furthermore, \cite{Kashino2016} find an anti-correlation between metallicity and SFR in FMR when using \cite{Mannucci2010} metallicity calibration; it retrieves to a positive trend when using \cite{Dopita2016} metallicity calibration. This is similar to our result that the trend is reversed when using N2O2 and N2S2 indexes metallicity calibrations.
    
    More recently, \cite{Curti2019} derived the MZR for a sample of SDSS galaxies with $T_{e}$ metallicity calibration, they found the turnover mass and the saturation metallicity are in agreement with previous MZR studies, while showing significantly lower normalization compared to those based on photoionization models.  This determination is similar to our result as showing in Figure \ref{fig:mass_z_sfr} and \ref{fig:plot_mz_all}. \cite{Cresci2019} also reanalyzed the data of CALIFA and SDSS-IV MaNGA samples, and found the secondary relation between MZR and SFR, which are different from the findings in \cite{Barrera-Ballesteros2017} and \cite{Sanchez2017}. \cite{Cresci2019} demonstrated that at fixed mass the metallicity depends on SFR, and shows an inversion at high stellar mass region caused by the metallicity calibration and SFR estimation, which is similar to our result. 

    In conclusion of this section, the different metallicity calibrations give different correlations between the mass surface density, metallicity and local star formation surface density (in Figures \ref{fig:mass_z_sfr} and \ref{fig:plot_mz_all}). These results may be caused by metallicity calibration, the $\Sigma_{\rm SFR}$ determination, and the ionization parameter. In general, the electronic temperature method, derived with the ratios of faint auroral to nebular emission lines (e.g., $\oiii\lambda4363 / \oiii\lambda5007$), is regarded as the most reliable approach for the metallicity estimation. However, it is not suitable for spectrum with no such an auroral line detection. To estimate the metallicity with these spectra, O3N2 and N2 diagnosis are calibrated by empirical fitting to the electronic temperature metallicity with emission line ratios \citep{Pettini2004}, while the N2O2 and N2S2 diagnosis are based on the photoionization models for $\hii$ regions \citep{Kewley2002,Dopita2016}. Because of the different approaches in the calibration, a large discrepancy is found between the metallicities derived by different calibrators \citep{Morisset2016}. For example, the O3N2 and N2 diagnosis can not accurately reproduce the supersolar metallicity, the N2O2 and N2S2 calibrators are model-dependent and then provide some uncertainty.  As we investigate in Figure \ref{fig:plot_ion_z}, we can see the strong correlation between the ratio of O32 and the local metallicity, this is known as the ionization parameter is sensitive to oxygen abundance. Determination of metallicity using strong emission lines may bias the ionization parameter, and it is necessary to find the accurate oxygen abundance. Due to this, there is a correlation between metallicity and local star formation rate surface density. Our result is partly consistent with reported negative correlation between metallicity and SFR {\citep{Mannucci2010,Andrews2013}, and the positive correlation in \cite{Kashino2016} at the higher mass bins.

\section{SUMMARY}
\label{sec:summary}
  This work aims to explore the dependence of local $\Sigma_{*}$ -- $Z$ -- $\Sigma_{\rm SFR}$ relation using the  740,000 spaxels from the MaNGA survey. We try to demonstrate if the local $\Sigma_{\rm SFR}$ exists or not in FMR. Here, we summarize our findings as follows.
  \begin{itemize}
    \item We present local $\Sigma_{*}$ -- $Z$ -- $\Sigma_{\rm SFR}$ relation using metallicity calibration of O3N2, N2, and N2O2 and N2S2. We find a strong correlation with local $\Sigma_{\rm SFR}$ using O3N2, N2, but the trend is not obvious with N2O2 and N2S2. This means that the existence of FMR probably depends on the metallicity calibrators. 
    
    \item We find a mixed relation of $\Sigma_{\rm SFR}$ -- $Z$ (anti-correlation/positive correlation) at lower/higher stellar mass bins according to metallicity derivation. In particular,  when using the metallicity from N2O2 and N2S2 indexes, the existence of FMR vanishes at higher masses, while the correlation retrieves using a similar metallicity scale with \cite{Mannucci2010}.
    \item We confirm the dependence of the ratio of O32 on metallicity, which means that a higher ionization parameter is found at lower metallicity, and the lower ionization parameter tends to higher metallicity.
    \item We provide the FMR after exclusion of low stellar mass, which is a projection of local stellar mass surface density, metallicity, and star formation rate surface density, and find that the scatters are reduced when using N2 and O3N2 indexes, comparing to the $\Sigma_*$ -- $Z$ relation.
  \end{itemize} 
    
\acknowledgments

This work is supported by the National Key R\&D Program of China (2017YFA0402600), the B-type Strategic Priority Program of the Chinese Academy of Sciences (XDB41000000) and the National Natural Science Foundation of China (NSFC, Nos. 11421303, and 11973038). Berzaf B. acknowledges support from CAS-TWAS President's Fellowship. We thank Enci Wang for useful discussion. Y.L.G. gratefully acknowledges support from the China Scholarship Council (No. 201906340095).

Funding for the Sloan Digital Sky Survey IV has been provided by the Alfred P. Sloan Foundation, the U.S. Department of Energy Office of Science, and the Participating Institutions. SDSS-IV acknowledges support and resources from the Center for High-Performance Computing at the University of Utah. The SDSS web site is www.sdss.org. SDSS-IV is managed by the Astrophysical Research Consortium for the Participating Institutions of the SDSS Collaboration including the Brazilian Participation Group, the Carnegie Institution for Science, Carnegie Mellon University, the Chilean Participation Group, the French Participation Group, Harvard-Smithsonian Center for Astrophysics, Instituto de Astrofsica de Canarias, The Johns Hopkins University, Kavli Institute for the Physics and Mathematics of the Universe (IPMU)/University of Tokyo, Lawrence Berkeley National Laboratory, Leibniz Institut fr Astrophysik Potsdam (AIP), Max-Planck-Institut fr Astronomie (MPIA Heidelberg), Max-Planck-Institut fr Astrophysik (MPA Garching), Max-Planck-Institut fr Extraterrestrische Physik (MPE), National Astronomcal Observatory of China, New Mexico State University, New York University, University of Notre Dame, Observatao Nacional/MCTI, The Ohio State University, Pennsylvania State University, Shanghai Astronomical Observatory, United Kingdom Participation Group, Universidad Nacional Autonoma de Mexico, University of Arizona, University of Colorado Boulder, University of Oxford, University of Portsmouth, University of Utah, University of Virginia, University of Washington, University of Wisconsin, Vanderbilt University, and Yale University.

\bibliography{ms}
\end{document}